\input{aipcheck}

\documentclass[
    ,final            
  ]
  {aipproc}

\layoutstyle{8x11single}

\usepackage{amsmath,amssymb}

\begin{document}

\title{Neutralino Dark Matter in MSSM Models with Non-Universal Higgs Masses}


\author{Pearl Sandick}{
  address={Theory Group and Texas Cosmology Center,\\
The University of Texas at Austin, TX 78712, USA}
}

\begin{abstract}

We consider the Minimal Supersymmetric Standard Model (MSSM) with varying amounts of non-universality 
in the soft supersymmetry-breaking contributions to the Higgs scalar masses.  In addition to the constrained MSSM 
(CMSSM) in which these are universal with the soft supersymmetry-breaking contributions to the squark and slepton 
masses at the input GUT scale, we consider scenarios in which both the Higgs masses are non-universal by the same 
amount (NUHM1), and scenarios in which they are independently non-universal (NUHM2).  As the lightest neutralino is a 
dark matter candidate, we demand that the relic density of neutralinos not be in conflict with measurements by WMAP and 
others, and examine the viable regions of parameter space.  Prospects for direct detection of 
neutralino dark matter via elastic scattering in these scenarios are discussed.

\end{abstract}

\maketitle


\section{Introduction}

TeV-scale Supersymmetry (SUSY) is one of the most compelling theories of physics beyond the Standard Model for 
several reasons: it facilitates unification of the gauge couplings, as expected in Grand Unified Theories (GUTs); 
it stabilizes the Higgs vacuum expectation value, offering a solution to the related heirarchy and naturalness 
problems of the Standard Model; and it predicts a light Higgs boson, as expected from electroweak precision 
measurements.  In addition, if R-parity is conserved, the lightest supersymmetric particle (LSP) is stable, and, 
if uncharged, is therefore a natural particle candidate for astrophysical cold dark matter.


Phenomenologically, one of the most-studied versions of the Minimal Supersymmetric Standard Model (MSSM) is 
the constrained MSSM (CMSSM), in which the soft supersymmetry-breaking contributions to the masses of the SUSY partners 
of the quarks and leptons and the Higgs scalars are universal at some GUT input scale, as motivated by minimal 
supergravity.  However, while 
the CMSSM may be the simplest scenario, it is by no means the most plausible version of the MSSM. Here, we present the
results of recent studies of models with non-universal supersymmetry-breaking contributions to the Higgs 
masses~\cite{eos08,eos09}.

The CMSSM is parametrized by four continuous parameters specified at the SUSY GUT scale and a sign choice: the 
universal gaugino mass, $m_{1/2}$, the universal scalar mass, $m_0$, the universal value for the trilinear couplings, 
$A_0$, the ratio of the Higgs vacuum expectation values, $\tan(\beta)$, and the sign of the Higgs mixing parameter, $\mu$.  In this 
scenario, the GUT-scale values of the effective Higgs masses are $m_1=m_2=m_0$, and the electroweak 
vacuum conditions fix $|\mu|$ and the pseudoscalar Higgs mass, $m_A$. In the NUHM1, the effective Higgs masses 
are assumed to be universal at the GUT scale, though the universality with $m_0$ is relaxed. One more input parameter, in addition to those of 
the CMSSM, 
is required, and  may be taken as $\mu$, $m_A$, or the GUT-scale value of the Higgs masses $m_1 = m_2$. 
Similarly, in NUHM2 models, no relation between $m_1$, $m_2$, and $m_0$ is assumed. This scenario may be parametrized 
by additional inputs $\mu$ and $m_A$, or by both GUT-scale effective Higgs masses. 

In each scenario, the renormalization group equations of the MSSM are used to determine the low-scale observables, and 
constraints from colliders and cosmology are applied.   We assume that the 
lightest neutralino is the dark matter candidate, calculate its abundance, and determine the prospects for direct detection for models 
which are phenomenologically and cosmologically viable, i.e. models not excluded by constraints from colliders, and where the upper limit on the relic density of neutralinos is respected. Direct 
detection experiments derive limits on the dark matter-nucleon scattering cross section under the assumption that the 
local density of cold dark matter comes from a single source, that is, one particle species is responsible for the 
entire dark matter abundance, $\Omega_{CDM}$.  However, this may not be the case. If the 
calculated neutralino abundance is less than the WMAP central value~\cite{wmap}, $\Omega_{CDM}$, we assume a secondary source of cold 
dark matter, and rescale the neutralino-nucleon cross sections by a factor $\Omega_{\chi} / \Omega_{CDM}$ in order to compare our calculated elastic scattering cross sections with limits from direct searches.

\section{The CMSSM}

\begin{figure}
  \includegraphics[height=.36\textheight]{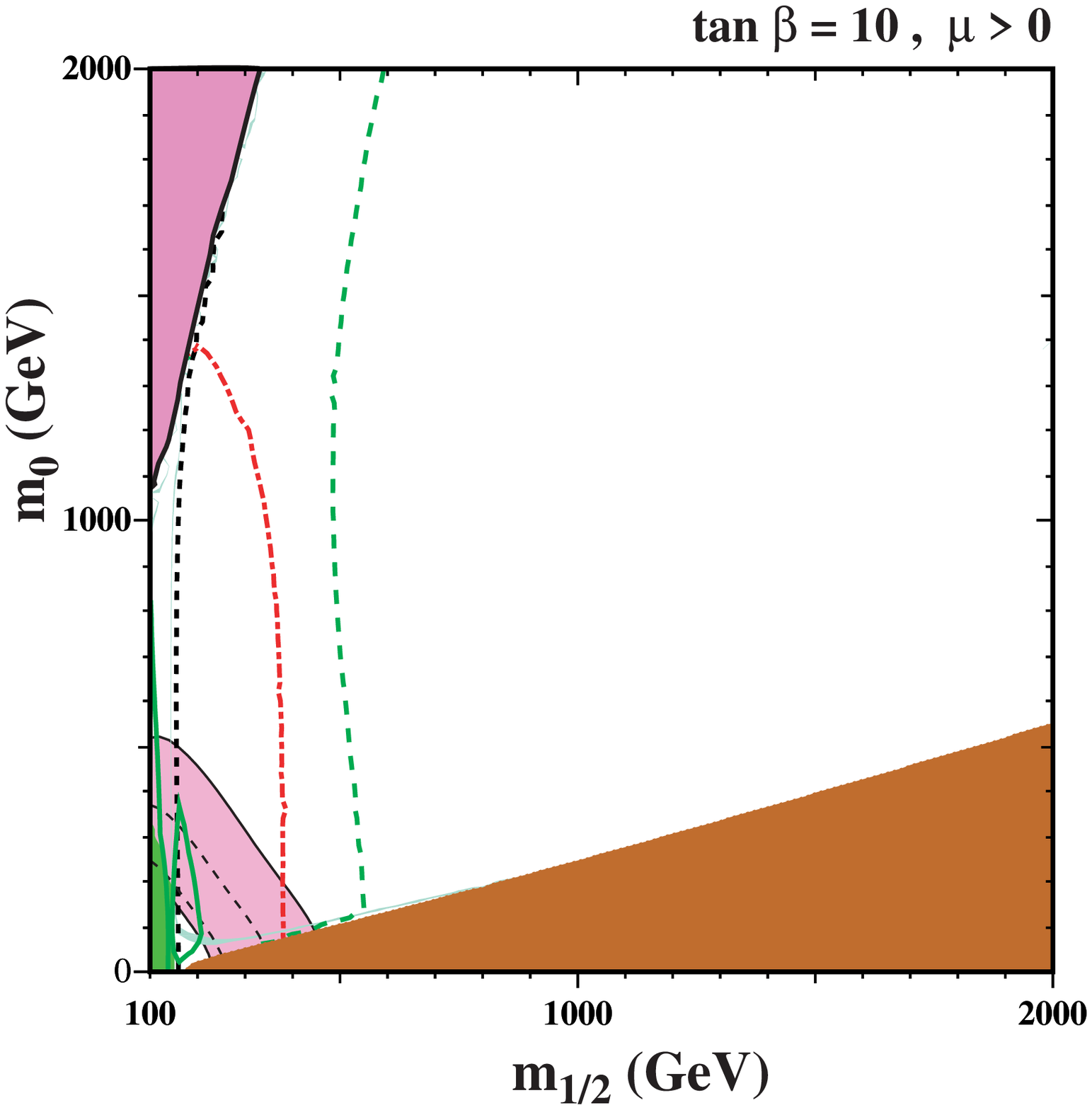}
  \includegraphics[height=.36\textheight]{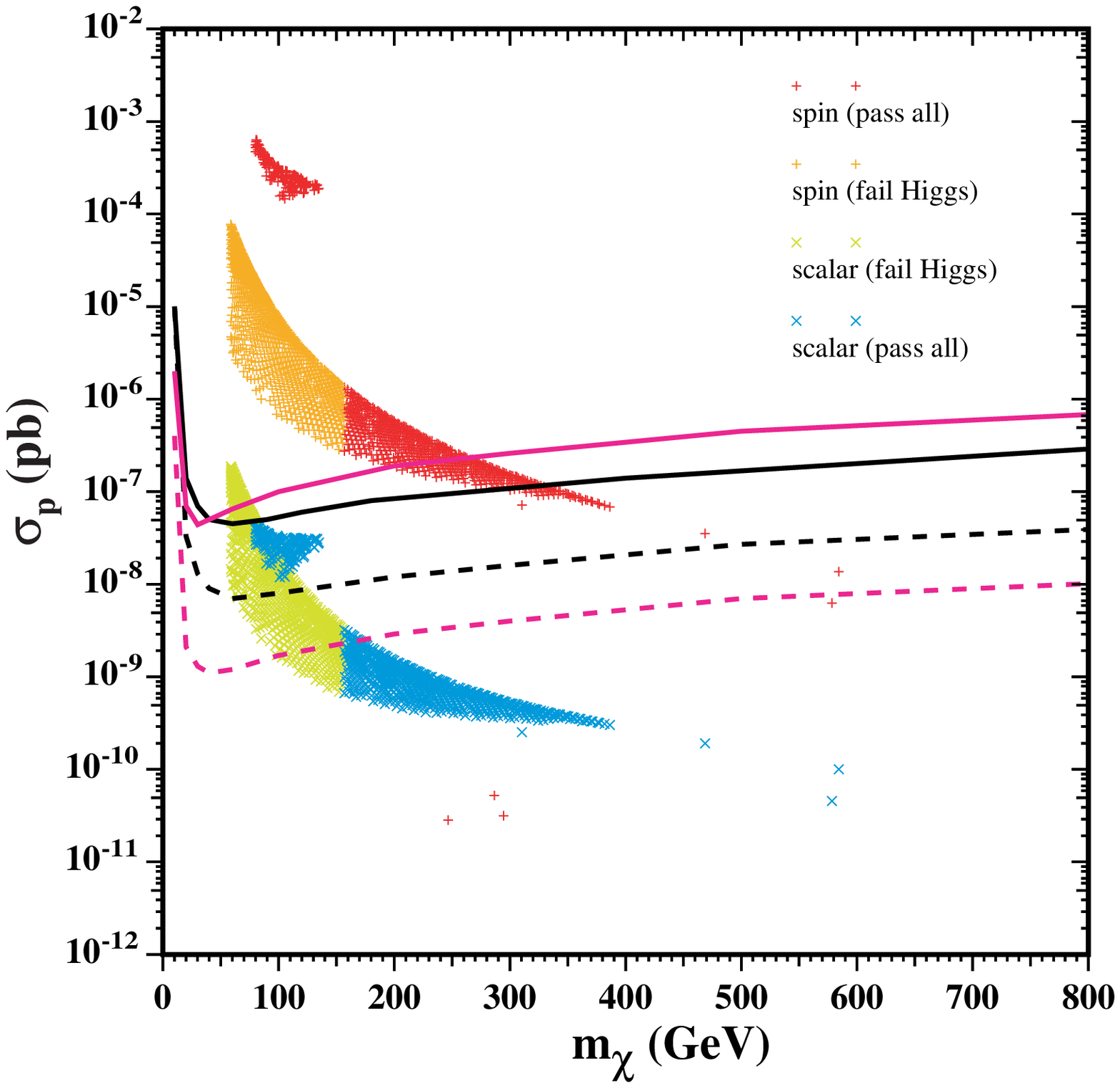}
  \caption{In the left panel, we show the CMSSM $(m_{1/2},m_0)$ plane for $\tan\beta=10$, $A_0=0$, and $\mu>0$ 
with the corresponding neutralino-nucleon elastic scattering cross sections as functions of neutralino mass in the right 
panel. Constraints are as discussed in the text. 
For further details on this and other figures, please refer to~\cite{eos09}. 
\label{fig:cmssmplane}}
\end{figure}

In Fig.~\ref{fig:cmssmplane}, we present the neutralino-nucleon elastic scattering cross section as a function of neutralino 
mass for one slice of the CMSSM parameter space. In the left panel, we show the $(m_{1/2},m_0)$ plane for $\tan\beta=10$, $A_0=0$, and $\mu>0$, with the 
corresponding neutralino-nucleon elastic scattering cross sections for phenomenologically-viable regions in the panel on the right.
The region excluded because the LSP is a charged $\widetilde{\tau}$ is shaded
brown, and, at large $m_0$, that where electroweak symmetry breaking cannot be obtained (resulting
in unphysical $\mu^2 < 0$), in pink. The red dot-dashed contour corresponds to a
Higgs mass of 114 GeV.  At lower $m_{1/2}$ the Higgs boson would be lighter, which is  
excluded by its non-observation at LEP~\cite{LEPsusy}.  We also plot a black dashed
contour for $m_{\chi^{\pm}}=104$ GeV, with the region at lower $m_{1/2}$ also
excluded by LEP.  The green shaded region at very low $m_{1/2}$ and $m_0$ is
disfavored by the measured branching ratio for $b \rightarrow s \gamma$~\cite{bsgex},
while the light pink shaded region is favored by the measurement of the muon
anomalous magnetic moment at the 2-$\sigma$ level~\cite{g-2}. Finally, in the turquoise
shaded regions, the relic density of neutralinos falls within the WMAP range~\cite{wmap}.
For the value $\tan\beta = 10$ presented here, the only cosmologically-preferred regions are the
coannihilation strip, bordering the $\widetilde{\tau}$-LSP region, and the focus-point region at large $m_0$,
where $\mu$ is small and the LSP is a mixed bino-Higgsino state.  Over the bulk of the
plane, the relic density of neutralinos exceeds the value measured by WMAP by more 
than 2-$\sigma$. There are, however, slim strips where $\Omega_{\chi} h^2$ is below the WMAP
range, which lie between the WMAP-preferred strips and the excluded regions they
border.

Turning to the right panel of Fig.~\ref{fig:cmssmplane}, we have separated out the cross 
section for  spin-dependent (SD) scattering (red/orange) and the scalar or spin-independent (SI) cross section (blue/green). Here 
and throughout, we assume the current limits on the branching ratios of $B \rightarrow \mu\mu$ and $b \rightarrow s \gamma$ are 
respected, 
as well as the lower limit on the chargino mass, and differentiate regions in which the scalar Higgs mass is below the 
LEP limit (green/orange) and above the LEP limit (blue/red). Here, the Higgs mass is below the LEP limit in the coannihilation strip at low $m_{1/2}$, corresponding to low $m_\chi$. Both the SI and SD cross sections in the upper right panel contain two distinct regions, that arising from the 
focus-point region
at $m_{\chi} \lesssim 150$ GeV and relatively large $\sigma$, and a more extended region due to the 
coannihilation strip.  In the coannihilation strip, 50 GeV $< m_{\chi} <$ 400 GeV, 
where the lower limit on $m_{\chi}$ is a result of the LEP constraint on the chargino mass,
and the upper limit on $m_{\chi}$ corresponds to the end-point of the coannihilation strip 
for $\tan \beta = 10$. At larger values of $\tan \beta$, we also find a rapid annihilation funnel in the CMSSM 
parameter space. 

We also plot in the right panel of Fig.~\ref{fig:cmssmplane} the limits on the 
SI cross section from CDMS~II~\cite{cdmsII} (solid black line)
and XENON10~\cite{XENON10} (solid pink line), as well as the sensitivities projected  
for XENON100~\cite{XENON100} (or a similar 100-kg liquid noble-gas detector such as 
LUX, dashed pink line) and SuperCDMS at the Soudan Mine~\cite{superCDMS} 
(dashed black line). One can see that the SD cross sections are generally larger than SI cross sections, though current upper limits are larger than $10^{-2}$ pb. 

At very low $m_{1/2}$, CDMS~II and XENON10 have definitively excluded some of the region 
where $m_h$ is below the LEP limit. We show this explicitly in the left panel
by plotting the reach of current and future direct detection experiments in the
parameter space.  Here and in subsequent parameter space scans, we display contours of
scalar neutralino-nucleon scattering cross section, scaled by $\Omega_{\chi} / \Omega_{CDM}$ if necessary,
of $5 \times 10^{-8}$ pb (solid green contours) and $10^{-9}$ pb (dashed green contours).
A cross section of $5 \times 10^{-8}$ pb is currently excluded by XENON10 for
$m_{\chi}=30$ GeV and by CDMS~II for $m_{\chi}=60$ GeV, and will be probed by
SuperCDMS for $m_{\chi}$ up to $\sim 1000$ GeV. Tonne-scale liquid noble-gas
detectors such as the proposed XENON1T or a similar detector mass for LUX/ZEP
will be sensitive to scalar cross sections below $10^{-9}$ pb for all neutralino masses
in the range 10 GeV $\lesssim m_{\chi} \lesssim$ a few TeV~\cite{XENON100}.  
These detectors  will be sensitive to cross sections below $10^{-10}$ pb over much of the preferred mass range $m_{\chi} \sim O(100)$ GeV.

\begin{figure}
  \includegraphics[height=.36\textheight]{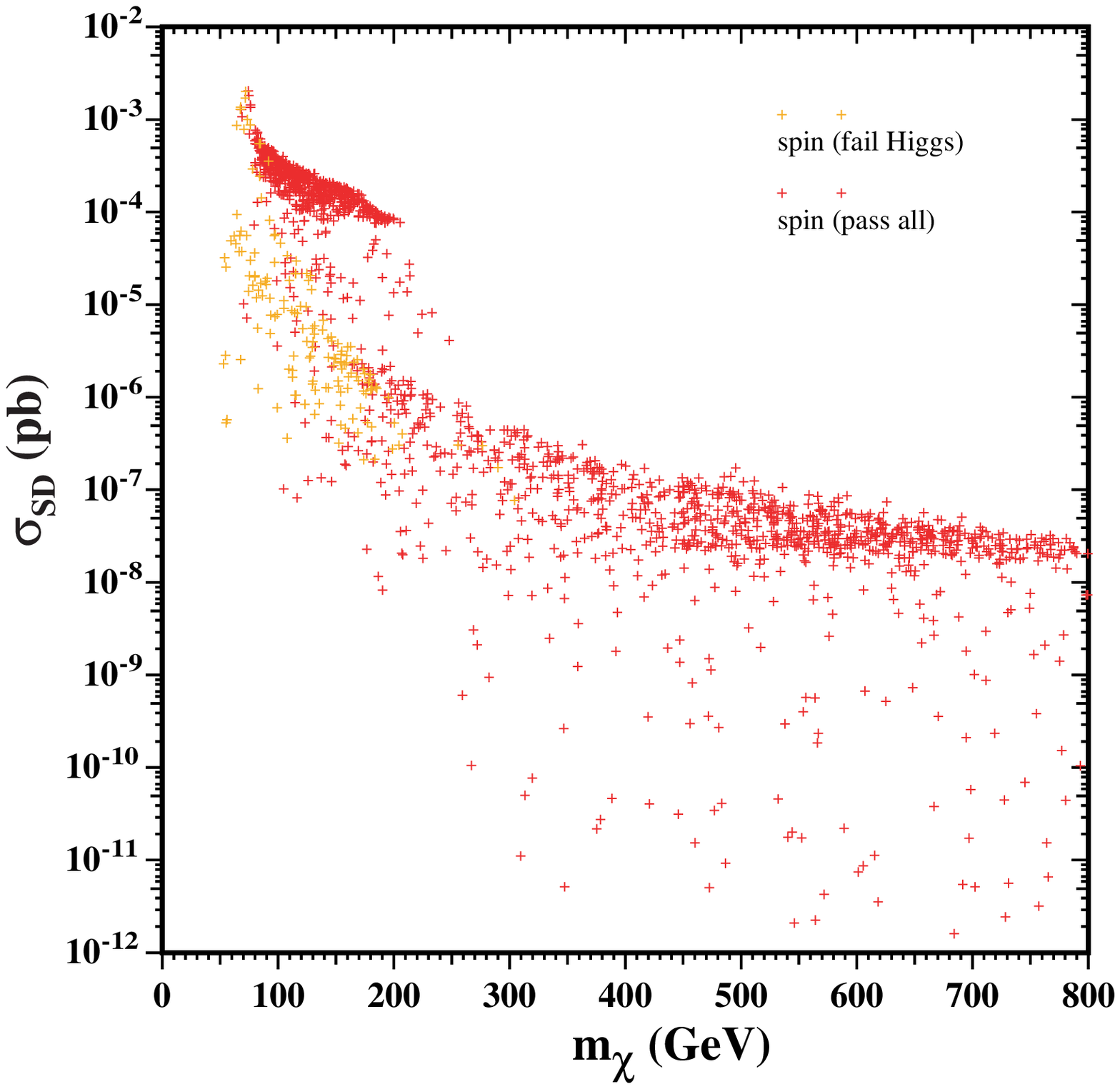}
  \includegraphics[height=.36\textheight]{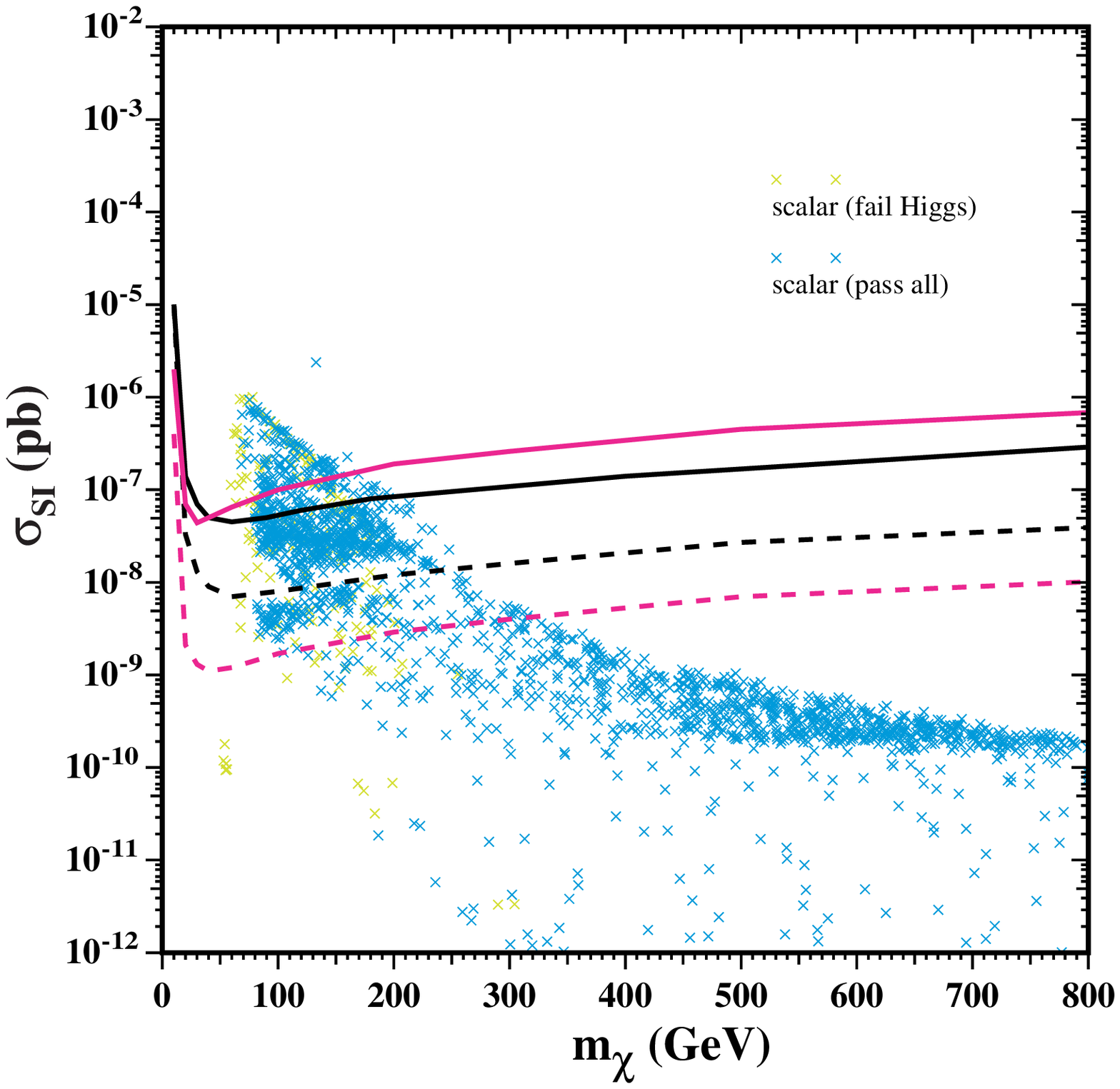}
\caption{Here we present the entire potential range of 
neutralino-nucleon elastic scattering cross sections as functions of neutralino mass for the CMSSM, with $5 \leq \tan\beta \leq 55$, 100 GeV $\leq 
m_{1/2} 
\leq 2000$ GeV, 0 $\leq m_0 \leq 2000$ GeV, and $-3 m_{1/2} \leq A_0 \leq 3 m_{1/2}$  We consider $\mu<0$ only for 
$\tan\beta<30$. On the left are the cross sections for SD scattering, and on the right are those for SI scattering. Also shown in the right 
panel are upper limits on the SI dark matter scattering cross section as specified in the text.
\label{fig:cmssmrand}}
\end{figure}

Of course, we would like to know the range of potential neutralino-nucleon elastic scattering cross sections that are possible in the CMSSM.  In Fig.~\ref{fig:cmssmrand}, we scan over all CMSSM parameters, with SD neutralino-nucleon elastic 
scattering cross sections in the left panel, and SI cross sections in the right panel. One can see that, in both 
the SI and SD cases, there is an 
upper limit on the cross section as a function of neutralino mass. Focusing on the SI elastic scattering cross 
sections, at low $m_{\chi} < 300$~GeV, cross sections generally 
exceed $
10^{-9}$~pb.  The largest cross sections, already excluded by CDMS and XENON10, come primarily from the focus point 
region at
 large $\tan \beta$.


\section{NUHM1 Models}

\begin{figure}
  \includegraphics[height=.36\textheight]{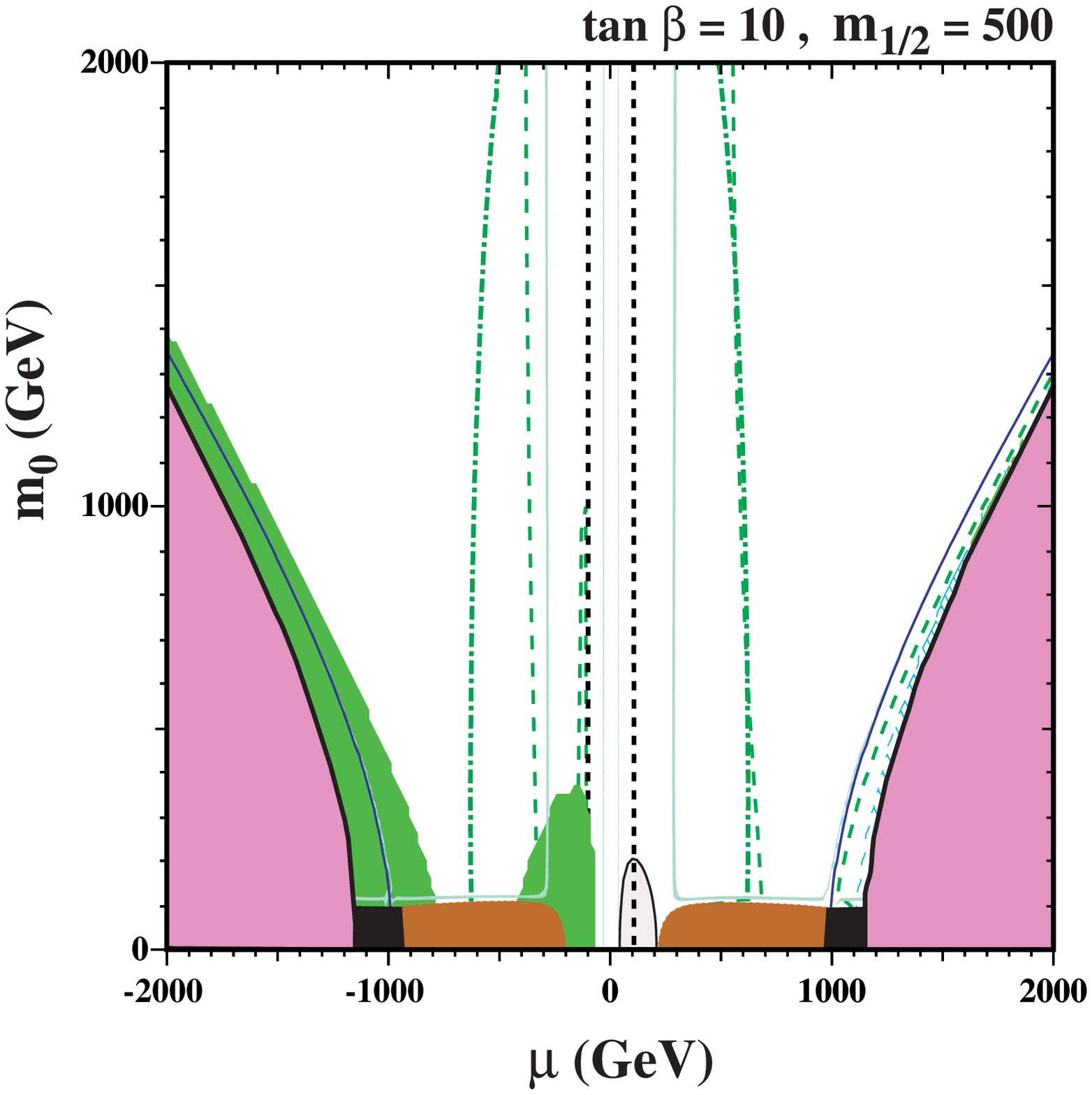}
  \includegraphics[height=.36\textheight]{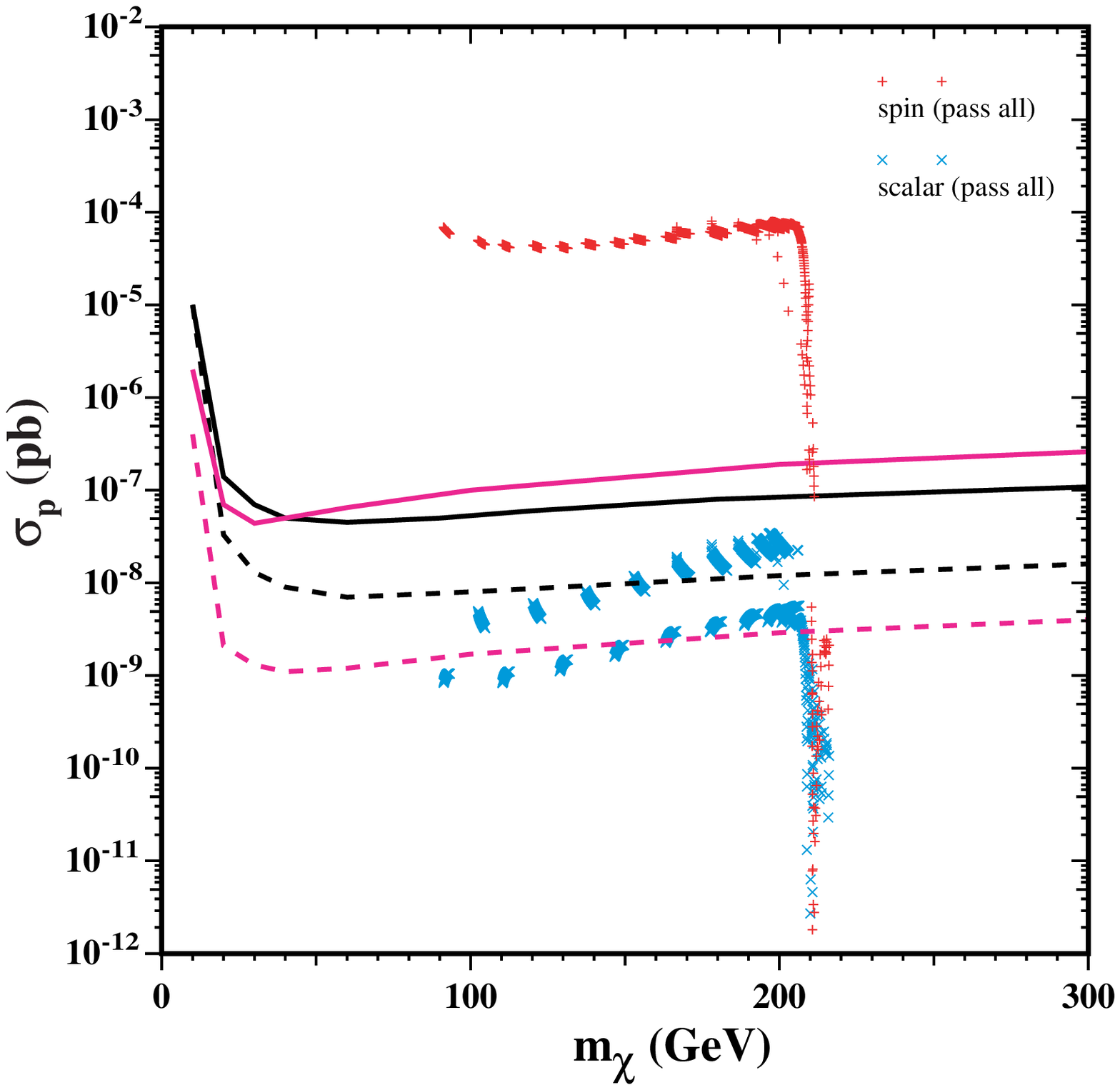}
  \caption{The left panel shows an NUHM1 $(\mu,m_0)$ plane for $m_{1/2}=500$ GeV, $\tan\beta=10$, and $A_0=0$. The right panel shows the corresponding neutralino-nucleon elastic scattering cross sections for phenomenologically-viable models.
\label{nuhm1muvm0}}
\end{figure}

In the NUHM1, the parameter space is expanded by one dimension, resulting in additional regions where all constraints are satisfied: {\it selectron} coahhihilation strips and {\it crossover} regions.  In the left panel of Fig.~\ref{nuhm1muvm0}, we show an example slice of 
NUHM1 parameter space, a $(\mu,m_0)$ plane. The cosmologically-preferred (turquoise) shaded strips, from large to small $|\mu|$, are: 
rapid annihilation funnels, where $m_\chi \approx m_A/2$, rising up at $|\mu| \approx 1000$ GeV and arcing outwards (note that for $\mu < 0$ the entire 
funnel is excluded by the branching ratio of $b \rightarrow s \gamma$); selectron coannihilation strips bordering the black excluded selectron-LSP regions at low $m_0$; stau coannihilation strips
 between $|\mu| \approx 1000$ GeV and $|\mu| \approx 300$ GeV at low $m_0$; and vertical crossover strips at $|\mu|\approx 300$ GeV, where the LSP is in a mixed bino-Higgsino state. 
The resulting neutralino-nucleon cross sections are shown in the right panel of Fig.~\ref{nuhm1muvm0}. Most cross sections are clustered 
at $m_\chi \sim 215$ GeV due to the fact that the lightest neutralino is bino-like over most of the plane.  The smallest cross sections shown come from the funnel or other regions where the relic density is suppressed (and therefore the cross section is scaled). The largest cross sections come from the crossover strip, near its intersection with the coannihilation strip, and for $\mu > 0$.  Between the crossover strips, moving to smaller $|\mu|$, $m_\chi$ decreases, as does the relic density of neutralinos, leading to strips in the right panel of scaled cross sections, with $m_\chi$ decreasing to $\approx 100$ GeV.
  
\begin{figure}
  \includegraphics[height=.36\textheight]{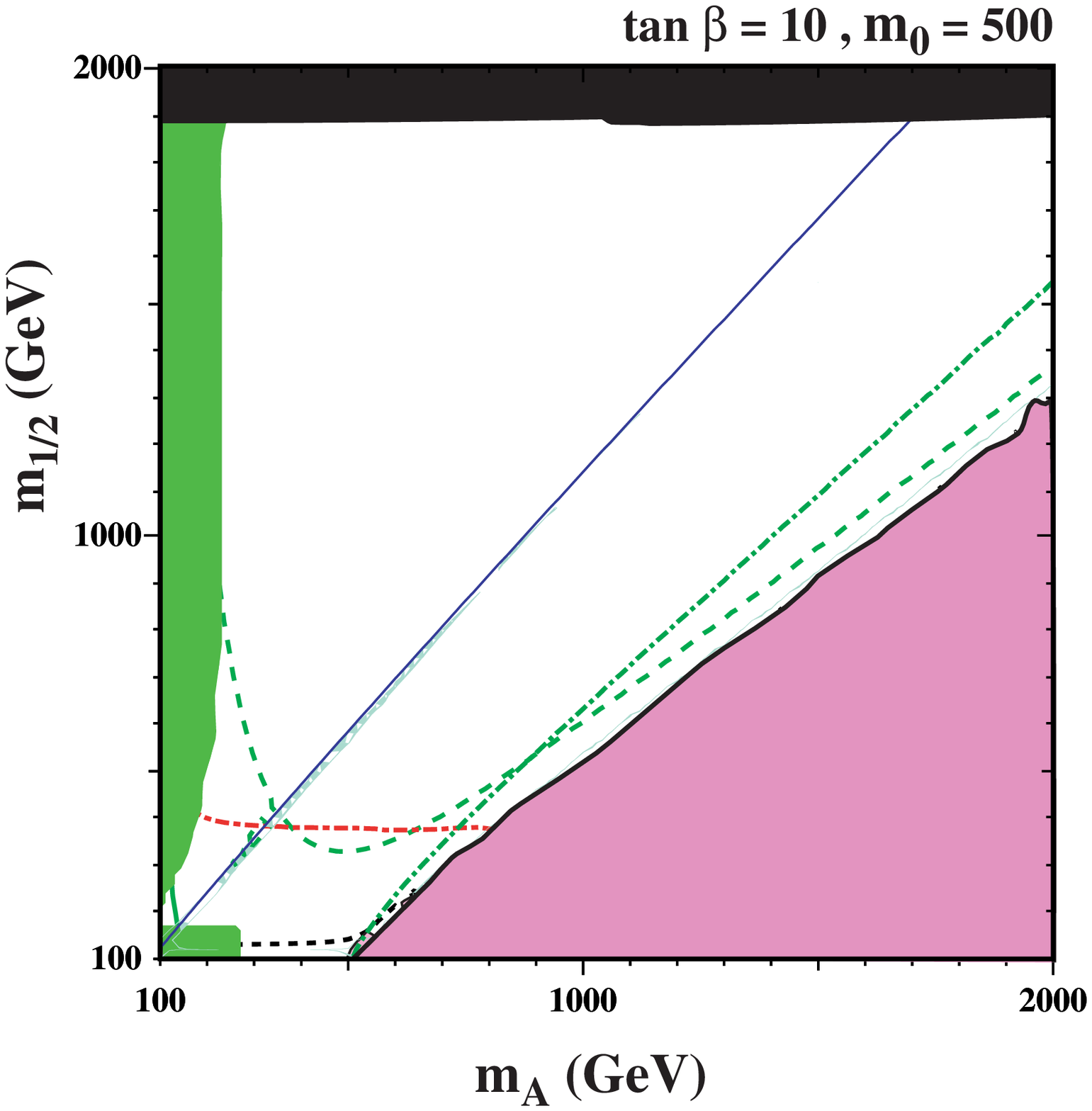}
  \includegraphics[height=.36\textheight]{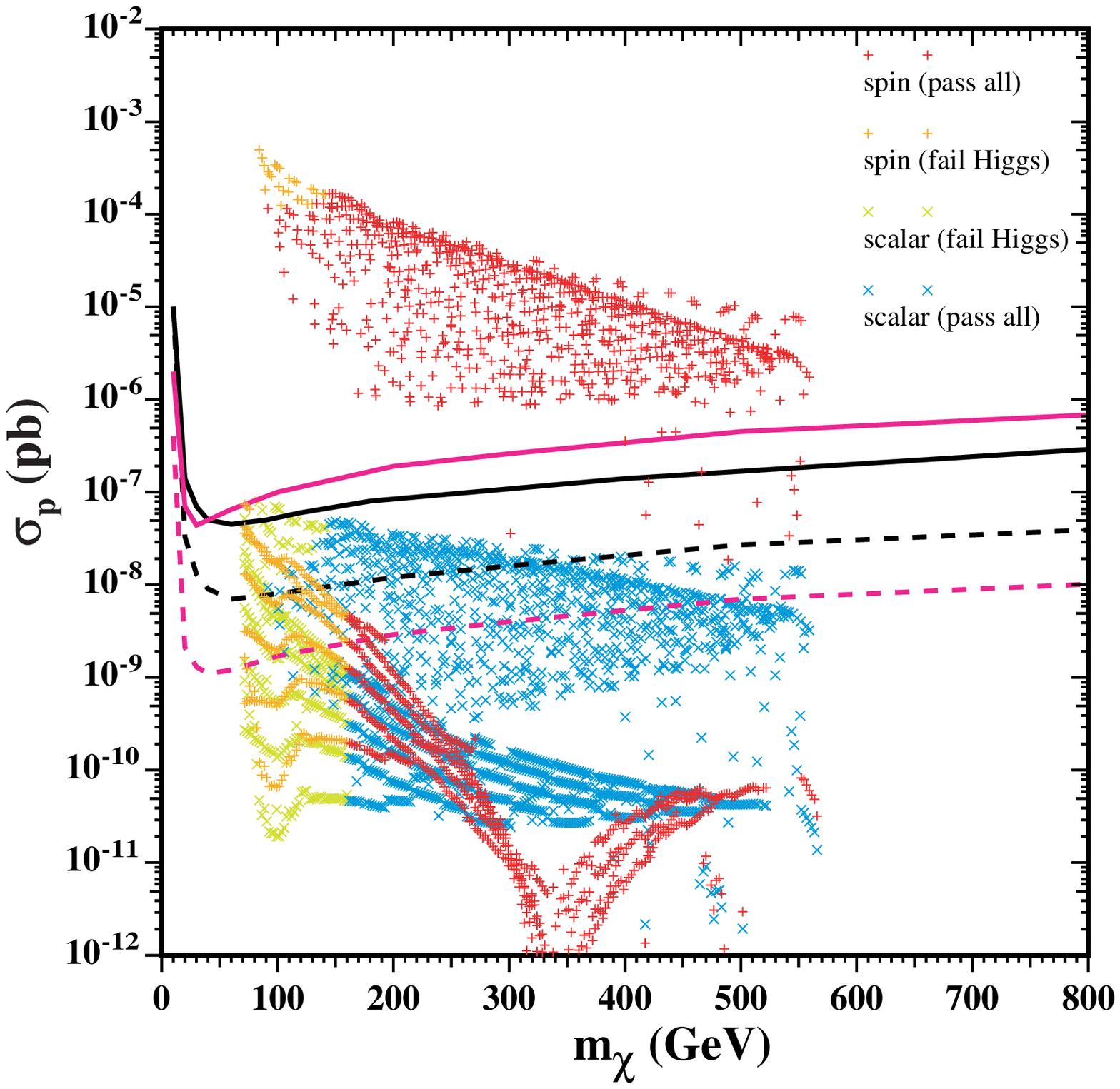}
  \caption{The left panel shows an NUHM1 $(m_A,m_{1/2})$ plane for $m_0=500$ GeV, $\tan\beta=10$, and $A_0=0$. The right panel shows the corresponding neutralino-nucleon elastic scattering cross sections for phenomenologically-viable models.
\label{fig:nuhm1mAvM}}
\end{figure}

In Fig.~\ref{fig:nuhm1mAvM}, we show a different slice of parameter space, this time the $(m_A,m_{1/2})$ plane, for $m_0=500$ GeV, $\tan\beta=10$, $\mu>0$, and $A_0=0$. The triangular allowed region is bounded by the constraint on the branching ratio of $b \to s \gamma$
at small $m_A$, the appearance of a slepton LSP at large $m_{1/2}$, and the absence of a
consistent electroweak vacuum at larger $m_A$ and smaller $m_{1/2}$.
The diagonal blue line indicates where $m_\chi = m_A/2$. On each side of this contour there is a 
narrow rapid-annihilation funnel strip where the relic density falls within the WMAP
range. The funnel extends only to $m_{1/2} \approx 1200$ GeV, and therefore
$m_{\chi} \lesssim 550$ GeV for this region. There is another WMAP strip in the
focus-point region close to the electroweak vacuum boundary,
where the LSP is more Higgsino-like. The displayed part of the focus-point 
strip is cut off at $m_A = 2000$ GeV, corresponding to $m_{\chi} \lesssim 600$ GeV: 
larger values of $m_\chi$ would be allowed if one considered larger values of $m_A$.
Apart from the region between this strip and
the boundary, and between the funnel strips, the relic LSP density would exceed the WMAP
range. The limit on the branching ratio of $b \rightarrow s \gamma$ is important at very 
low $m_A$ and $m_0$, but it is the constraint on the Higgs mass (red
dot-dashed curve, roughly horizontal at $m_{1/2} = 400$ GeV)
that places a lower limit on the expected LSP mass
for the funnel region of $m_\chi \sim 160$ GeV.

The green dashed contour in the left panel of Fig.~\ref{fig:nuhm1mAvM}, indicating a future sensitivity to
spin-independent dark matter scattering at the level of $10^{-9}$ picobarns, runs mainly 
through the region where the relic density exceeds the WMAP upper limit.
SuperCDMS at Soudan would be sensitive to much of the focus-point region
shown in the panel on the left, but a more advanced detector would be required 
if the relic density
of neutralinos is below the WMAP range. Unfortunately, even if neutralinos
make up all the dark matter in the universe, much of the funnel region will remain out of
reach, even to next-generation direct detection experiments.
Points associated with the funnel with cross sections larger than $10^{-9}$ pb
fail the Higgs mass constraint.

The right panel of Fig.~\ref{fig:nuhm1mAvM} displays the scattering cross sections
for phenomenologically-viable points in the left plane. The fact that scalar cross sections in the funnel region are
generally smaller than those in the focus-point region is reflected in the lower cutoff on
the LSP mass, $m_{\chi} \lesssim 550$ GeV, seen in the right panel for the points with
$\sigma_{SI} \lesssim$ few $\times 10^{-10}$ pb.  In the focus-point region,
the LSP is in a mixed state with $m_{\chi} \lesssim 600$ GeV, and the cross section may be
much larger: by two orders of magnitude for the SI cross section,
and four orders of magnitude for the SD cross section. These large
cross sections at large $m_\chi$ have no counterparts in the CMSSM, where mixed states in the focus-point region are much lighter. Furthermore,
in the CMSSM, the focus point is reached only at large $m_0$, whereas here it appears even at fixed $m_0 = 500$ GeV.
CDMS~II and XENON10 are already beginning to probe he large scalar cross sections in the focus-point region, with a few points at very low $m_{1/2}$
already being excluded, as one can see directly in the right panel\footnote{We note
these points also have $m_h < 114$~GeV and are within the solid green contour in the left panel of Fig.~\ref{fig:nuhm1mAvM}
at very low values of $m_{1/2}$ and $m_A$.}. We also note that there are points in
the right panel with very low cross sections even though they have $m_\chi < 150$~GeV,
which also have no counterparts in the CMSSM. These points are associated with the
funnel and their cross sections are scaled down due to the low
relic density in that region.

A new feature seen in this plot is the near-vanishing of the spin-dependent
cross section when $m_\chi = 350$ GeV.  This feature
is associated with the funnel region where elastic scattering cross sections are substantially lower than those of the focus-point region.  However, near $m_\chi = 350$ GeV there is the possibility  
for a complete cancellation in the SD neutralino-nucleon elastic scattering matrix elements when the spin
contribution from up quarks cancels that due to down and strange quarks. The exact position of the
cancellation is sensitive to the values of the spin matrix elements adopted,
but the existence of the cancellation is robust. Of course, as we present here the cross sections for scattering on protons, the
cross sections for scattering on neutrons will not exhibit a cancellation in the same place.


\section{NUHM2 Models}

As already discussed, the NUHM2 has two parameters in addition to those
already present in the CMSSM, which may be taken as free choices
of both the quantities $m_A$ and $\mu$. As the number of parameters is relatively large, a systematic survey of the NUHM2 parameter space is quite complex. Here we exhibit one plane of parameter space whose
features we compare with the CMSSM and NUHM1.  For further discussions of direct detection cross sections in the NUHM2,
see \cite{eos09,EFlOSo}.

\begin{figure}
  \includegraphics[height=.36\textheight]{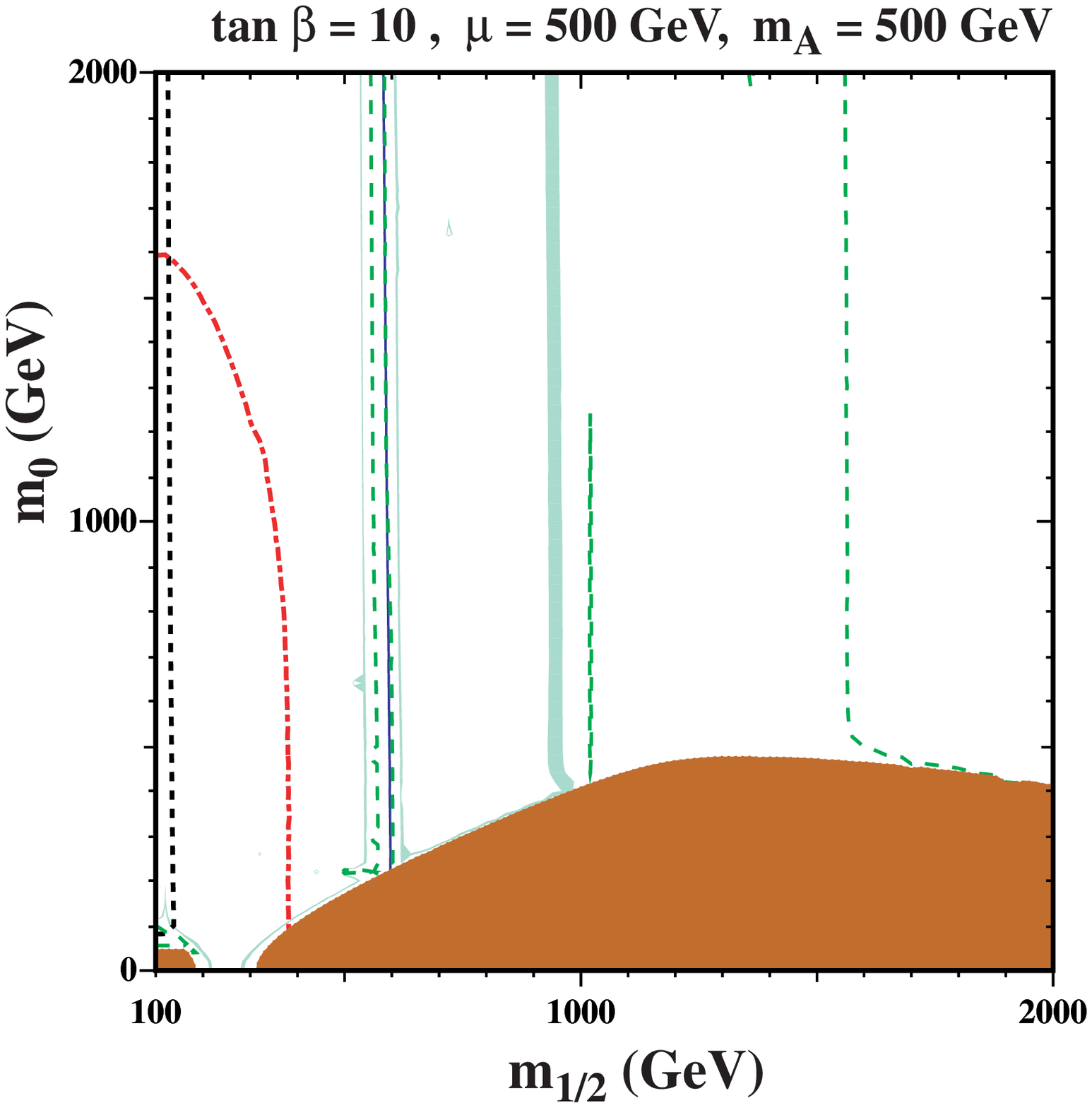}
  \includegraphics[height=.36\textheight]{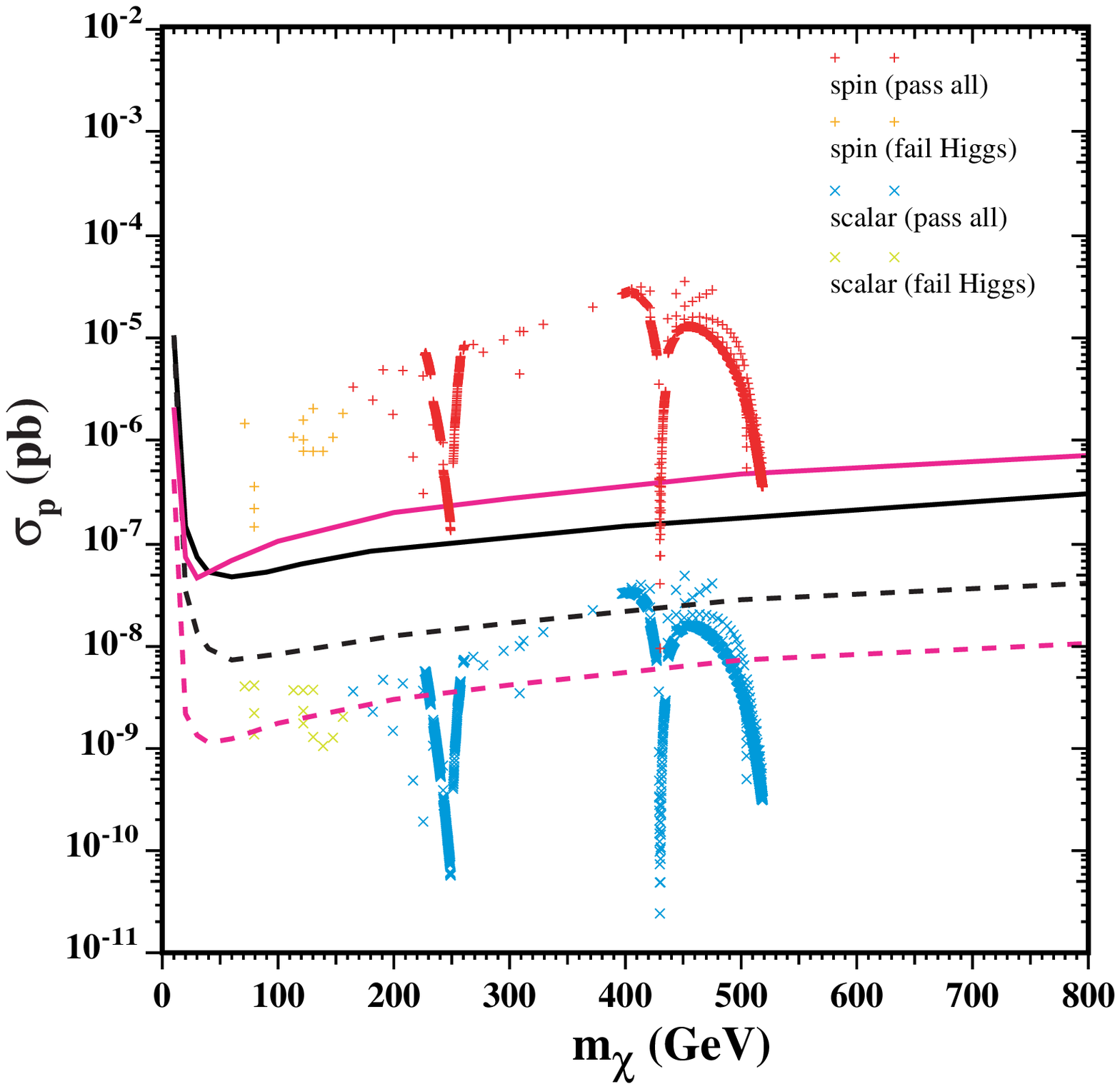}
  \caption{The left panel shows an NUHM2 $(m_{1/2},m_0)$ plane for $\mu=m_A=500$ GeV, $\tan\beta=10$, and $A_0=0$. The right panel shows the corresponding neutralino-nucleon elastic scattering cross sections for phenomenologically-viable models.
\label{fig:nuhm2}}
\end{figure}

We display in the left panel of Fig.~\ref{fig:nuhm2} a sample NUHM2 $(m_{1/2}, m_0)$ plane
with $\tan\beta = 10$ and fixed $m_A = 500$~GeV and $\mu = 500$~GeV,
which reveals a few interesting new features. As in the CMSSM, there is a region in the plane
at large $m_{1/2}$ and small $m_0$ which is forbidden because the lighter stau
is the LSP. Just above this forbidden region, as in the CMSSM, there is a
stau-coannihilation strip. However, jutting up from this strip at $m_{1/2} \sim 600$~GeV
and $\sim 950$~GeV, there are vertical strips where the relic density of neutralinos falls within
the WMAP range. The double strips at $m_{1/2} \sim 600$~GeV form a rapid-annihilation
funnel on either side of the (solid blue) contour where $m_A = 2 m_\chi$.
Again, such funnels appear only at large $\tan\beta$ in the CMSSM, but the freedom to choose
different values of $m_A$ in the NUHM2 permits the appearance of a rapid-annihilation
funnel also at the low value $\tan\beta = 10$ shown here. The other vertical WMAP strip
appears because, as $m_{1/2}$ increases relative to $\mu$ (which is fixed here),
the Higgsino fraction
of the lightest neutralino increases, which in turn increases the
annihilation rate, decreasing the relic density.  In this case, this results in
a crossover region when $m_{1/2} \sim 950$~GeV.
At slightly larger $m_{1/2} = 1020$ GeV, the lightest and next-to-lightest neutralinos are nearly degenerate in mass, and rapid coannihilations (through $Z$-exchange) result in a narrow region
with suppressed relic density, and therefore a suppressed scalar cross section.

These novel regions are clearly visible in the right panel of Fig.~\ref{fig:nuhm2}.
The elastic scattering cross sections do not vary rapidly as $m_\chi$ increases, until
the funnel region at $m_\chi \sim 250$~GeV is reached. The "V"-shaped suppression 
in the cross section arises from the increasing value of $m_0$ as one rises up the funnel.
The two sides of the funnel approach each other as $m_0$ increases, eventually
joining together and resulting in a minimum value where the two sides of the funnel meet (at a value of $m_0 > 2000$~GeV). After this excitement, the cross
section continues to rise gradually as one follows the coannihilation strip, until the
crossover strip is reached at $m_\chi \sim 400$~GeV. Here the cross section decreases
again as $m_0$ increases, to values even smaller than in the rapid-annihilation funnel,
before rising again and finally declining towards the end of the coannihilation strip.
Note that the entire region to the right of the transition strip is viable, albeit with
a relic density below the WMAP range.  The cross section is therefore reduced
due to scaling.  Because the neutralino is predominantly Higgsino-like here,
its mass is given by $\mu$ rather than $m_{1/2}$, and so we see in the right panel that the largest values of $m_\chi$ correspond to our choice of fixed $\mu$.

\section{Summary}

\begin{figure}
  \includegraphics[height=.36\textheight]{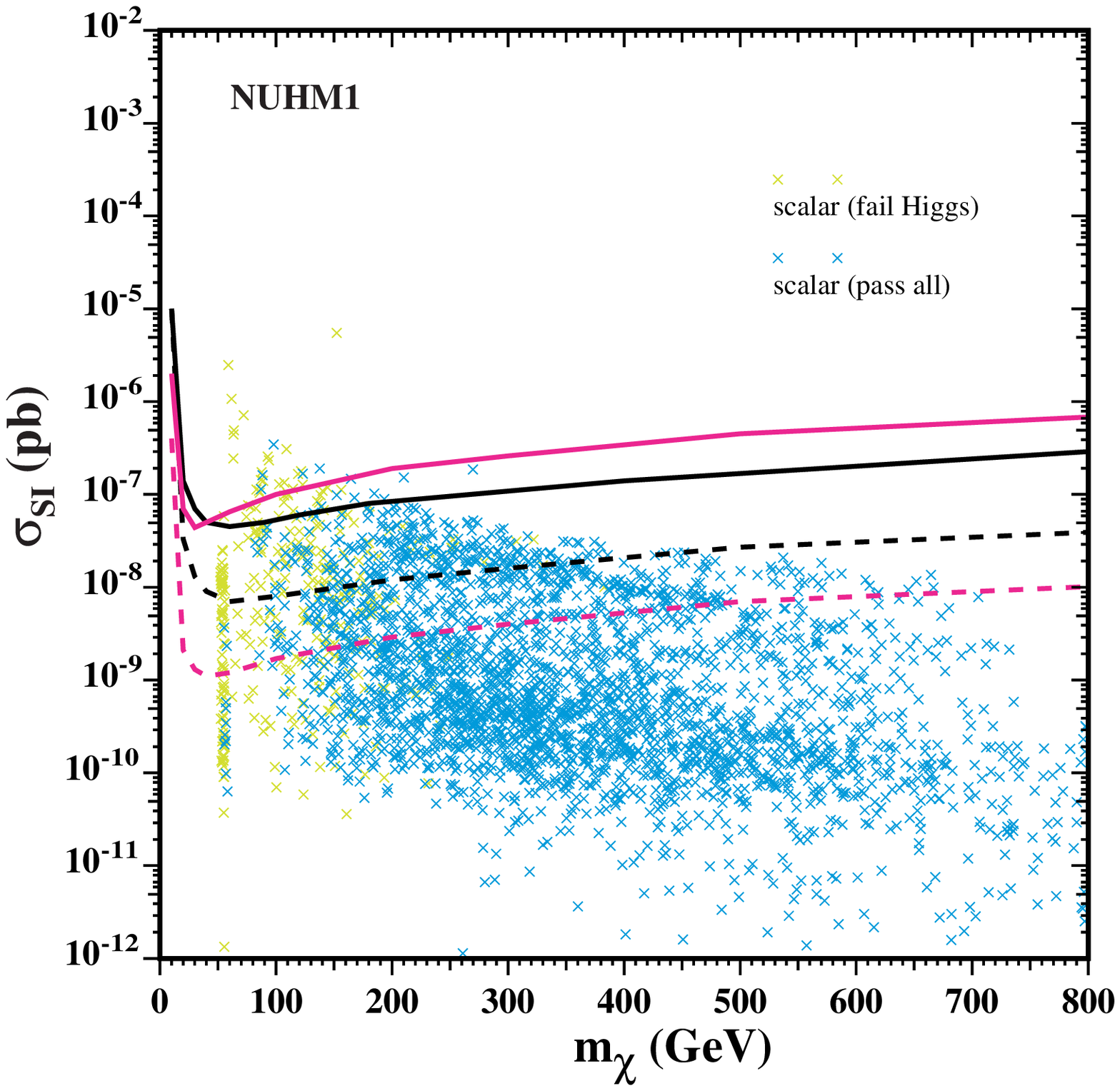}
  \includegraphics[height=.36\textheight]{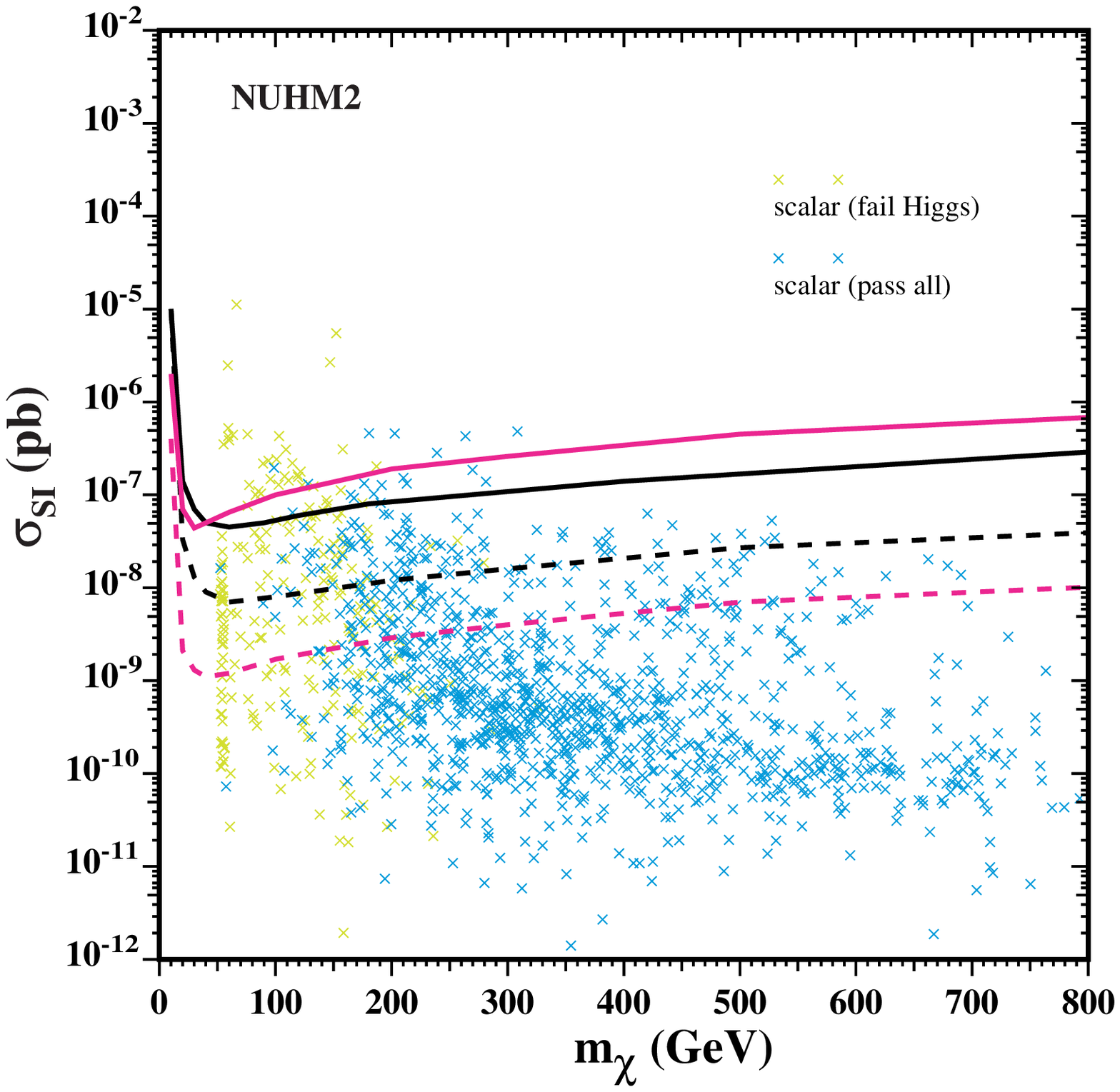}
  \caption{We show the entire potential ranges in the NUHM1 and NUHM2 of 
the SI neutralino-nucleon cross sections with respect to neutralino mass.  In both plots, we scan $5 \leq \tan\beta 
\leq 55$, 100 GeV $\leq m_{1/2} \leq 2000$ GeV, 0 $\leq m_0 \leq 2000$ GeV, and
$-3 m_{1/2} \leq A_0 \leq 3 m_{1/2}$. The GUT-scale 
values of $m_1$ and $m_2$ are in the range $(-2000,2000)$ GeV, with $m_1 = m_2$ in the NUHM1. 
\label{nuhmrand}}
\end{figure}

Finally, in Fig.~\ref{nuhmrand} we show the entire potential ranges for the SI neutralino-nucleon elastic scattering cross section as a function of neutralino mass in the NUHM1 (left) and NUHM2 (right). Because of the additional freedom in the Higgs 
sector, and since the SI elastic scattering cross sections receive important contributions from Higgs exchange, there is more 
variability in the cross sections in the NUHM1 and NUHM2 than in the CMSSM. This is clear from comparison with the corresponding CMSSM panel in Fig.~\ref{fig:cmssmrand}. Differences are due primarily to the appearance of additional regions 
of parameter space in NUHM models where the lightest neutralino is in a mixed bino-Higgsino state, such as the crossover region.

In both the NUHM1 and NUHM2, we find significantly larger cross sections at larger $m_\chi$ than would be expected in the CMSSM.  These cross sections may be probed by the next generation of direct detection experiments for $m_\chi$ as
large as $\sim 700$ GeV, while in the CMSSM there is no hope for detection with these instruments for $m_\chi \gtrsim 350$ GeV.  In addition, we find viable points at very low $m_\chi$ with rather low cross sections. If nature has given us a neutralino of this character, then we may not directly detect dark matter for some time, but the LHC may find supersymmetric particles fairly quickly. In either scenario, we will know that the CMSSM is not an adequate description of nature. We look forward to results from both collider experiments and the next generation of dark matter direct detection experiments, which may give us some interesting hints about the mechanism of supersymmetry breaking.


\begin{theacknowledgments}
This material is based upon work supported by the National Science Foundation under Grant No. PHY-0455649.
\end{theacknowledgments}

\end{document}